# A Proposal for a Nationwide Student Gradebook Information Network

Cristina TURCU, Cornel TURCU, Evelina GRAUR

University of Suceava, 13 University Street, 720229-Suceava, Romania

{cristina, cturcu, evelyn}@eed.usv.ro

**Abstract**

Today's students are encouraged to study and develop expertise in more than one national academic environment. As a matter of fact, their educational activities inevitably occur in a variety of academic settings and even span several years. Consequently students' academic results and progress are expected to be easily monitored and accessed nationally. The authors of the present paper have devised a student gradebook information network to be nationally employed by all public and private universities and colleges. The papers deals with the architectural principles underlying the system and discusses aspects related to data collection, data analysis and data storage across multiple machines while providing a seamless view of entire infrastructure and service delivery system from a single Web access point. The utility of the architectural system is discussed in relationship with its major advantages: user-friendliness, security access, flexibility, transparency, distributional power, and scalability. The major beneficiary of the system is the Romanian higher education system.

## 1. Introduction

Currently, Internet technologies are widely used all over the world. Almost every aspect of our daily life is influenced and often improved by the use of the WWW. The educational environment is at the same time the initiator and the beneficiary of various Intranet-Internet systems conceived to assist a wide range of activities in areas such as learning, instruction, evaluation, administration and management. One particular area of interest that has recently come to our attention is the mobility of students within the same university or among higher education institutions.

In most European countries the education system comprises a network of institutions distributed all over the national territory and generally characterized by the heterogeneity and diversity of their structure and organization, logistics, instructional methods and resources, and even cultural context.

Although every individual institution displays a well established hierarchical structure, they often require the integration of specialized functional levels with their own needs and features. As a consequence, one may distinguish two major, yet slightly conflictual, directions in dealing with national information networks: on the one hand such networks are summoned to preserve the individuality of their institutional users manifested through their specific needs and features; on the other hand, these specificities must find their proper place in the overall structure, both locally and nationally.

There are numerous applications already installed and running in most educational institutions and, certainly, one cannot underestimate their capacity of meeting the specific needs of their users. Hence our major concern is to develop and propose the most suitable technical solution to integrate all existing applications and guarantee their inter-operationality. Besides acknowledging all previously made investments, our solution must keep the sub-systems operational while their monolithic structures are being altered and prepared to accept modular developments. Ultimately, the financial aspect of this large-scale project is not something to be ignored.

## 2. The current state of affairs

At the moment there is practically no Romanian university to lack a system of managing student information. The capabilities and features of these applications have been tailored in accordance with the degree of complexity of the institutional level they have been summoned to assist. In all Romanian faculties there is one general procedure related to handling gradebooks: the secretary's office issues the gradebook in print format which is handed over to the teacher on the examination day; the teacher fills in the gradebook and then returns the gradebook report to the secretary's office to be processed. The secretary introduces the grades into the local information system and stores the data for its future usage in transcripts and student progress reports. This circuit is visually represented in Figure 1. As one can easily notice, the existing applications permit the local administration and structuring of all student-related information. Upon faculty or department request, these systems can automatically generate individual or group reports.

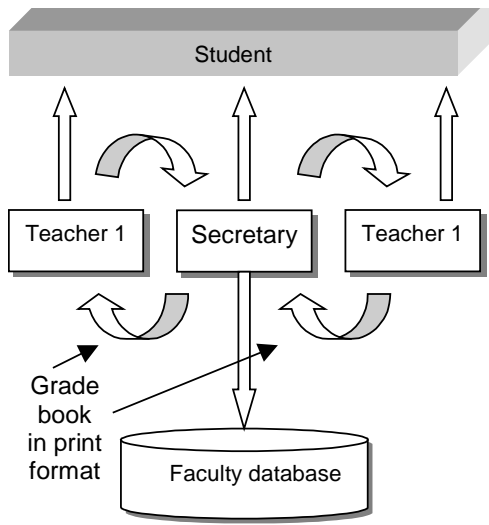

Figure 1. The general path of the Romanian academic grading task

However, as we have already hinted at, these applications already running in most faculties and departments are not interrelated; the systems do not communicate with each another. Only a national information network would be able to take into consideration the informative implications of students' mobility within the same university, their concomitant participation in other university-based study programs or their transfer requests.

When students decide to get transferred to other institutions, their academic records must be submitted in written format. At the moment there is no unitary application in Romania to allow the storage of student information, both personal and academic. Furthermore, there is no template to store all data. Most applications allowing this procedure have been locally developed and thus no well-defined design criteria have been applied. Under these circumstances, student registration following a transfer request is complicated and time-consuming. Even if the transcripts or other student progress reports are submitted in written format, they often require verification and validation. Besides ensuring a speedy access to all stored data, a national information network would be more attractive both technically and financially.

## 3. The architecture of the network

The architecture of the national network allows the collection, integration and availability of educational information at all points of the local network. The proposed architecture is presented in figure 2.

Within our network, the data coming from various sources will be administered by the faculty where the student has been initially registered. To eliminate all redundancies, the data will be stored in a central database called **Student Data Central Repository** (SDCR). All queries addressed by various departments about one and the same student are directed to the SDCR; thus the desired information can be obtained from a common repository. The operations involving data entering and editing are properly secured so that only authorized staff can access them. Moreover, special security measures are envisaged for the whole application.

Within the Nationwide Student Gradebook Information Network, student identification may be achieved by means of a unique student identification number such as the **Personal Numeric Code.** In case another identification procedure is desired, one must nationally agree upon and then clearly stipulate the assigning rules of the selected coding system.

An authorized user may access the SDCR database in various ways:

- by accessing the internal university network: either from a personal computer which requires the entering of a username and password, or by a personal RFID card (tag). The latter possibility is to be favoured because, apart from the rapid access to the database, it also allows the user to store various data on his/her RFID card.
- WAP (Wireless Application Protocol)
- PDA (Personal Digital Assistant)

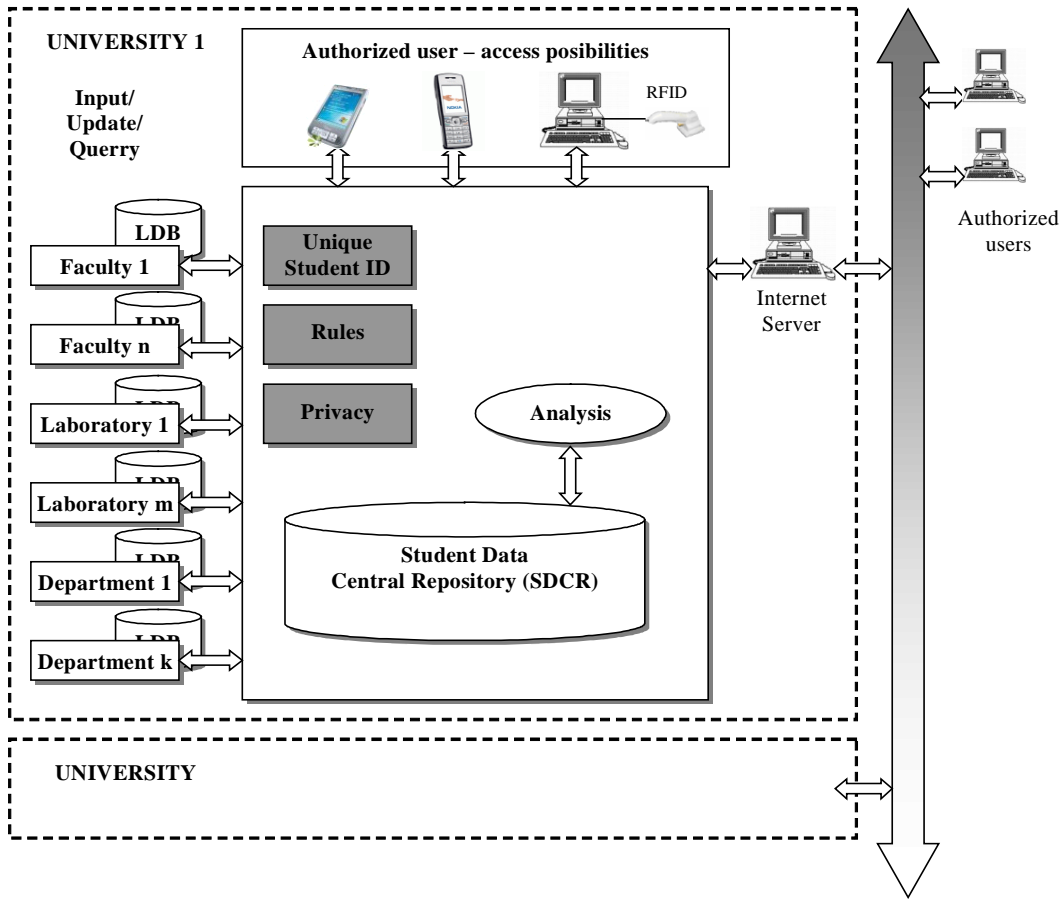

Figure 2. The architecture of the proposed network

Users are also permitted to access the SDCR database from personal computers outside the university network. Even other categories of users can be identified and authorized to access only certain informative fields in the database.

The accessing of data in real time presupposes either the implementation of an electronic gradebook system at the national level or the inter-relation of the already running local applications within a national network. In this way, any authorized user will have a unified picture on any given student.

The exchange of academic data is strongly related to the ontologies employed. Various solutions can be mentioned here:

- the usage of a common ontology by all universities (however this aspect presupposes the change of the overall memory data format, which is not cost effective)

- The architecture does not propose globally agreed ontologies; rather educational institutions reconcile their semantic differences through a mediator component.

The mediator component uses ontologies based on a standard as references to facilitate semantic mediation among involved institutions. Mediators can have a P2P communication architecture to provide scalability and to facilitate the discovery of other mediators. The overall architecture of the system could be adapted from the Semantic Web enabled Web services proposal [1];

- The use of intelligent agents allowing the integration, collection and analysis of all student-related information [2, 3].

## 4. Advantages

Our proposed network has the advantage of showing the progress of individual student enrolled in any academic program and the easy transfer of all student grades and scores in accordance with the specific curricula adopted by every national universities (both state and private) or by various faculties within the same university. Categories of authorized users are able to see student data in real time. The increased visibility of students' academic progress may enhance immediate feedback, which is likely to modify student behaviour. As timely feedback from the web-based gradebooks helps to modify student behaviour, student performance is also expected to rise.Moreover, since all data is centralized, a wide range of student reports and analyses can be prepared.

## 5. Conclusions

Although there is no higher education institution to lack a student information management system, the implemented systems have not been designed to communicate with each another. The mobility of students within the same university or across national universities would benefit from the creation of a national information network. Interconnected in real-time with the Electronic Gradebook Record System in every university, the Nationwide Student Gradebook Information Network lays the foundation of an academic information infrastructure. Mentioned should be made, however, that the proposed network is not intended to force the introduction of a unique equalizing model for the educational and organizational activities typical of every university, but to propose a common set of educational services to be accessed at any given moment by various categories of authorized users. At a later stage, in view of today's globalizing tendencies, the system can be further developed to allow the exchange of information with worldwide universities.